\documentclass{sig-alternate-05-2015}

\usepackage{xspace}
\usepackage[utf8]{inputenc}
\usepackage[T1]{fontenc}
\usepackage{microtype}
\usepackage{threeparttable}

\usepackage[table]{xcolor} 
\usepackage{subfig}
\usepackage{url}

\usepackage{booktabs}
\usepackage{tabularx}
\newcolumntype{R}{>{\raggedleft\arraybackslash}X}

\usepackage{tikz}
\usepackage{tikz-qtree}

\usepackage{dblfloatfix} 

\usepackage{todonotes}
\presetkeys%
    {todonotes}%
    {inline,caption={}}{}%

\newcommand{\ISO}[0]{ISO\slash IEC~9126\xspace}
\newcommand{\shortISO}[0]{ISO$^a$\xspace}

\CopyrightYear{2016} 
\setcopyright{acmcopyright}
\conferenceinfo{ICSE '16,}{May 14-22, 2016, Austin, TX, USA}
\isbn{978-1-4503-3900-1/16/05}\acmPrice{\$15.00}
\doi{http://dx.doi.org/10.1145/2884781.2884788}

\begin{document}
\title{Are ``Non-functional'' Requirements really Non-functional?}
\subtitle{An Investigation of Non-functional Requirements in Practice}

\numberofauthors{1}
\author{
\alignauthor
Jonas Eckhardt, Andreas Vogelsang, and Daniel M\'endez Fern\'andez\\
       \affaddr{Technische Universit\"at M\"unchen, Germany}\\
       \email{\{eckharjo|vogelsan|mendezfe\}@in.tum.de}
}

\maketitle

\begin{abstract}

Non-functional requirements (NFRs) are commonly distinguished from functional requirements by differentiating {\itshape how} the system shall do something in contrast to {\itshape what} the system shall do. This distinction is not only prevalent in research, but also influences how requirements are handled in practice. NFRs are usually documented separately from functional requirements, without quantitative measures, and with relatively vague descriptions.
As a result, they remain difficult to analyze and test.
Several authors argue, however, that many so-called NFRs actually describe behavioral properties and may be treated the same way as functional requirements. In this paper, we empirically investigate this point of view and aim to increase our understanding on the nature of NFRs addressing system properties. We report on the classification of 530 NFRs extracted from 11 industrial requirements specifications and analyze to which extent these NFRs describe system behavior.
Our results suggest that most ``non-functional'' requirements are {\itshape not} non-functional as they describe behavior of a system. Consequently, we argue that many so-called NFRs can be handled similarly to functional requirements.
\end{abstract}

\category{D.2.1}{Software Engineering}{Requirements\slash Specifications}

\terms{Experimentation, Documentation, Measurement}

\keywords{Non-functional requirements, classification, model-based development, empirical studies}

\section{Introduction}
One conventional distinction between non-functional requirements (NFRs) and functional requirements is made by differentiating {\itshape how} the system shall do something in contrast to {\itshape what} the system shall do~\cite{robertson2012mastering,Sommerville97}. This distinction is not only prevalent in research, but it also influences how requirements are elicited, documented, and validated in practice~\cite{Ameller12,borg2003bad,chung1995dealing,svensson2009quality}. 
As a matter of fact, up until now there does not exist a commonly accepted approach for the NFR-specific elicitation, documentation, and analysis~\cite{borg2003bad,svensson2009quality}; NFRs are usually described vaguely~\cite{borg2003bad,Ameller12}, remain often not quantified~\cite{svensson2009quality}, and as a result remain difficult to analyze and test~\cite{Ameller12,borg2003bad,svensson2009quality}. Furthermore, NFRs are often retrofitted in the development process or pursued in parallel with, but separately from, functional requirements~\cite{chung1995dealing} and, thus, are implicitly managed with little or no consequence analysis~\cite{svensson2009quality}. This limited focus on NFRs can result in the long run in high maintenance costs~\cite{svensson2009quality}.

Although the importance of NFRs for software and systems development is widely accepted, the discourse about how to handle NFRs is still dominated by how to differentiate them exactly from functional requirements~\cite{Broy16Rethinking,glinz2007non}. One point of view is that the distinction is an artificial one and we should rather differentiate between {\itshape behavior} (e.g., response times) and {\itshape representation} (e.g., programming languages). The underlying argument is that most NFRs actually describe behavioral properties~\cite{glinz2007non} and should be treated the same way as functional requirements in the software development process~\cite{Broy15NFR}. Behavioral properties subsume classical functional requirements, such as {\itshape ``the user must be able to remove articles from the shopping basket''} as well as NFRs which describe behavior such as {\itshape ``the system must react on every input within 10ms''}. Representational properties include NFRs that determine how a system shall be syntactically or technically represented, such as {\itshape ``the software must be implemented in the programming language Java''}~\cite{Broy15NFR,Broy16Rethinking}.

In this paper, we empirically investigate this point of view and aim to increase our understanding on the nature of NFRs addressing system properties. To this end, we classify 530 NFRs extracted from 11 industrial requirements specifications with respect to their kind. Our results show that 75\% of the requirements labeled as ``non-functional'' in the considered industrial specifications describe system behavior and only 25\% describe the representation of the system. As behavior has many facets, we further classify behavioral NFRs according to the {\itshape system view} they address (interface, architecture, or state), and the {\itshape behavior theory} used to express them (syntactic, logical, probabilistic, or timed)~\cite{Broy15NFR,Broy16Rethinking}. Based on this fine-grained classification, we discuss the implications we see on handling NFRs in the software engineering disciplines, e.g., testing or design.

Based on the results of our study, we conclude that most ``non-functional'' requirements are misleadingly declared as such because they actually describe behavior of the system. This in turn means that many so-called NFRs can be handled similarly to functional requirements. Please note that the focus of this paper is not to criticize the term ``non-functional'' but to expose the artificial separation of functional and non-functional (or quality) requirements in practice.

The remainder of the paper is structured as follows: In Section~\ref{bg}, we discuss background and related work, and, subsequently, we present our study design in Section~\ref{studydesign}. We report on the results in Section~\ref{results} and discuss the threats to validity and our mitigation strategies in Section~\ref{threats}. In Section~\ref{discussion}, we provide a discussion of the overall results and their impact on theory and practice, before concluding our work in Section~\ref{conclusion}.

\section{Background \& Related Work}
\label{bg}
In this section, we provide background and related work on NFR classifications and on the implications of NFRs on software development.

{\bfseries \itshape Previously published material.} In our previously published paper~\cite{Eckhardt15}, we presented a research proposal with the goal of analyzing natural language NFRs taken from industrial requirements specifications to better understand their nature. Our study reported here, relies on and extends our previous study design. We present the results in full detail, and provide a comprehensive discussion on the implications on software engineering disciplines.

{\bfseries \itshape NFR classifications.}
There exist several classification schemes for NFRs in literature (e.g.,~\cite{chung2009non,glinz2007non,iso9126,Pohl2010RE,Sommerville97,Wagner12}). 
One example for such a classification, which is based on a quality model, is the \ISO~\cite{iso9126}. It defines external and internal quality of a software system and derives several quality characteristics (e.g., {\itshape Functionality--Security} or {\itshape Portability--Installability}). Sommerville further provides a classification scheme based on a distinction between {\itshape process requirements}, {\itshape product requirements}, and {\itshape external requirements}~\cite{Sommerville97}. We base our distinction of NFR classes on the \ISO classification. Furthermore, we exclude process requirements from our study, as they do not describe properties of the system itself. 

Pohl~\cite{Pohl2010RE} discusses the misleading use of the term ``non-functional'' and argues to use ``quality requirements'' for product-related NFRs that are not constraints.
Glinz~\cite{glinz2007non} performs a comprehensive review on the existing definitions of NFRs, analyzes problems with these definitions, and proposes a definition on his own. He highlights three different problems with the current definitions: a definition problem, i.e., NFR definitions have discrepancies in the used terminology and concepts, a classification problem, i.e., the definitions provide very different sub-classifications of NFRs, and finally a representation problem, i.e., the notion of NFRs is representation-dependent. In our study, we faced all of the three problems: we motivate our study based on the definition and classification problem and during the execution of our study, we faced the representation problem (see also our discussion on threats to validity in Section~\ref{threats}).
Although we agree on the critique about the obsolete and misleading notion of the term ``non-functional'', it still dominates the way requirements are handled in practice, as reflected in our data.

Mairiza et al.~\cite{mairiza2010investigation} perform a literature review on NFRs, investigating the notion of NFRs in the software engineering literature to increase the understanding of this complex and multifaceted phenomenon. Amongst others, they found about 114 different NFR classes. As a result of a frequency analysis, they found that the five most frequently mentioned NFR classes in literature are 
{\itshape performance}, {\itshape reliability}, {\itshape usability}, {\itshape security}, and {\itshape maintainability} (in that order). 
In our study, we got similar results: we found that the five most frequently used NFR classes in our industrial specifications are {\itshape security}, {\itshape reliability}, {\itshape usability}, {\itshape efficiency}, and {\itshape portability} (in that order).\footnote{We excluded {\itshape functionality} from this list, as it is not a classical NFR {\itshape class}.} While Mairiza et al. performed their analysis on available literature, our study analyzes NFRs documented in industrial projects.

{\bfseries \itshape NFRs and their implications on software development.}
One of the first studies that analyzed how to systematically deal with NFRs in software development was conducted by Chung and Nixon~\cite{chung1995dealing}. They argue that NFRs are often retrofitted in the development process or pursued in parallel with, but separately from, functional design and that an ad-hoc development process often makes it hard to detect defects early. They perform three experimental studies on how well a given framework~\cite{Mylopoulos92} can be used to systematically deal with NFRs. Svensson~et~al.~\cite{svensson2009quality} perform an interview study on how quality requirements are used in practice. Based on their interviews, they found that there is no NFR-specific elicitation, documentation, and analysis, that NFRs are often not quantified and, thus, difficult to test, and that there is only an implicit management of NFRs with little or no consequence analysis. Furthermore, they found that at the project level, NFRs are not taken into consideration during product planning (and are thereby not included as hard requirements in the projects) and they conclude that the realization of NFRs is a reactive rather than proactive effort. Borg et al.~\cite{borg2003bad} analyze via interviews how NFRs are handled in practice by the example of two Swedish software development organizations. They found that NFRs are difficult to elicit because of a focus on functional requirements, they are often described vaguely, are often not sufficiently considered and prioritized, and they are sometimes even ignored. Furthermore, they state that most types of NFRs are difficult to test properly due to their nature, and when expressed in non-measurable terms, testing is time-consuming or even impossible. Ameller~et~al.~\cite{Ameller12} perform an empirical study based on interviews around the question {\itshape How do software architects deal with NFRs in practice?} They found that NFRs were not often documented, and even when documented, the documentation was not always precise and usually became desynchronized. Furthermore, they state that NFRs were claimed to be mostly satisfied at the end of the project although just a few classes were validated. With respect to model-driven development, Ameller et al.~\cite{ameller2010dealing} show that most model-driven development (MDD) approaches focus only on functional requirements and do not integrate NFRs into the MDD process. They further identify challenges to overcome in order to integrate NFRs in the MDD process effectively. Their challenges include {\itshape modeling of NFRs at the PIM-level}, which includes the question {\itshape which types of NFRs are most relevant to the MDD process?} According to Ameller et al.~\cite{ameller2010dealing}, the few MDD approaches that support the modeling of NFRs can be classified into approaches that use UML extensions~\cite{fatwanto2008analysis,wada2010model,zhu2009model} or a specific metamodel~\cite{gonczy2009model,kugele2008optimizing,molina2009integrating} to model NFRs. In all of the approaches, functional requirements and NFRs are modeled separately. Damm et al.~\cite{damm2005boosting} suggest to overcome this separation and propose a so-called rich component model based on UML that integrates functional and NFRs in a common model. Similar approaches exist for specific classes of NFRs (e.g., for availability~\cite{Junker12}). The results of our study provide empirical support for the claim that NFRs and functional requirements are not very different with respect to behavior characteristics and, therefore, can be integrated in a common system model. 

All these studies highlight, so far, that NFRs are not integrated in the software development process and furthermore that several problems are evident with NFRs. In this paper, we use these problems as motivation and analyze what NFR classes can be found in practice and discuss how they can be integrated in the software development process.

\section{Study Design}
\label{studydesign}
In this section, we describe our overall goal, our research questions, and the design of our study.
\subsection{Goal and Research Questions}
The goal of this study 
is to increase our understanding on the nature of NFRs addressing system properties\footnote{In our study, we exclude those NFRs going beyond system properties, e.g., process requirements.}. In particular, we are interested in understanding to which extent these NFRs and their respective classes (e.g., security or reliability) describe system behavior and what kind of behavior they address. This allows us to discuss the implications on handling NFRs in the software engineering disciplines (e.g., testing or design).

To achieve our goal, we formulate the following research questions (RQs), which we cluster in two categories:
\paragraph{1. Distribution of NFR classes in practice} 
We examine the distribution of NFR classes in practice via two research questions:
\begin{itemize}
\item[\textbf{RQ1}] \textbf{What NFR classes are documented in practice?} 
With this RQ,  we want to get an overview of the NFR classes that are documented in practice.
\item[\textbf{RQ2}] \textbf{What NFR classes are documented in different application domains?} 
Under this RQ, we analyze whether there is an observable difference between the application domains w.r.t. the documented NFR classes. 
\end{itemize}

\paragraph{2. Nature of the NFR classes} 
We analyze the NFR classes with respect to their nature (behavioral or representational) and their kind of behavior via three research questions:
\begin{itemize}
\item[\textbf{RQ3}] {\bfseries How many NFRs describe system behavior?} 
With this RQ, we want to better understand how many NFRs describe system behavior and how many describe the representation of a system (\emph{behavioral} vs. \emph{representational}) and whether this varies for different NFR classes.
\item[\textbf{RQ4}] {\bfseries Which \emph{system views} do behavioral NFRs address?} 
With this RQ, we want to better understand the relation between NFR classes and the system (modeling) views that the NFRs address, e.g., {\itshape interface}, {\itshape architecture}, or {\itshape state behavior}.
\item[\textbf{RQ5}] {\bfseries In which type of \emph{behavior theory} are behavioral NFRs expressed?} 
With this RQ, we want to better understand the relation between NFR classes and behavior theories used to express the NFRs, e.g., logical, timed, or probabilistic description.
\end{itemize}

\subsection{Study Object}
The study objects used to answer our research questions constitute 11 industrial specifications from 5 different companies for different application domains and of different sizes with 346 NFRs\footnote{In the data preparation phase, we split non-singular NFRs into singular NFRs. Thus, the final number of analyzed NFRs is 530.} in total. We collected all those requirements that were explicitly labeled as ``non-functional'', ``quality'', or any specific quality attribute. The specifications further differ in the level of abstraction, detail, and completeness. We cannot give detailed information about the individual NFRs or the projects. Yet, in Table~\ref{tbl:studyobjects}, we summarize our study objects, their application domain, and show exemplary (anonymized) NFRs as far as possible within the limits of existing non-disclosure agreements.

\begin{table*}
\scriptsize
\caption{Overview of the study objects}
\label{tbl:studyobjects}
\renewcommand{\arraystretch}{1.3}
\begin{threeparttable}[b]
\centering
\begin{tabularx}{\textwidth}{llrrrX}\toprule
{\bfseries Spec.}&{\bfseries Application Domain\tnote{1}}&{\bfseries \# Reqs}&{\bfseries \# NFRs}&{\bfseries \% NFRs}&{\bfseries Exemplary NFR (anonymized due to confidentiality)}\\ \midrule
S1& BIS (Finance) &200&61&30.5\%&{\itshape The availability shall not be less than [x]\%. That is the current value.}\\
S2& BIS (Automotive) &177&40&22.6\%&{\itshape An online help function must be available. 
}\\
S3& BIS (Finance) &107&5&4.7\%&{\itshape  The maximal number of users that are at the same time active in the system is [x].}\\
S4& ES/BIS  (Travel Mgmt.) &38&14&36.8\% & {\itshape The [system] must run on [the operating system] OS/2.}\\
S5& ES/BIS (Travel Mgmt.) &69&16&23.2\% & {\itshape It must be possible to completely restore a running configuration when the system crashes.}\\
S6& ES (Railway) &35&14&40.0\% & {\itshape The delay between passing a [message] and decoding of the first loop message shall be $\leq$ [x] seconds.}\\
S7& ES  (Railway) &122&19&15.6\% &{\itshape The collection, interpretation, accuracy, and allocation of data relating to the railway network shall be undertaken to a quality level commensurate with the SIL [x] allocation to the [system] equipment.
}\\
S8& ES/BIS  (Traffic Mgmt.)&554& 128&23.1\% & {\itshape Critical software components shall be restarted upon failure when feasible.}\\
S9& ES (Railway)  &393& 12&3.0\% & {\itshape The [system] will have a Mean Time Between Wrong Side Failure (MTBWSF) greater than [x] h respectively a THR less than [x]/h due to the use of [a specific] platform.} \\
S10& ES (Railway) &122& 31&25.4\% & {\itshape The [system] system shall handle a maximum of [x] trains per line.}\\
S11& BIS (Facility Mgmt.) &24&6&25.0\% & {\itshape The architecture as well as the programming has to guarantee an easy and efficient maintainability.}\\
\hline
$\Sigma$ 11& & $\Sigma$ 1.841&$\Sigma$ 346&18.8\%\\
\bottomrule
\end{tabularx}
\begin{tablenotes}
            \item [1] We distinguish BIS (Business Information Systems), ES (Embedded Systems), and hybrids of both. For reasons of simplicity, we group various domains (e.g. ``Finance'') according to single family of systems and use the term ``application domain'' for that classification.
        \end{tablenotes}
\end{threeparttable}
\end{table*}

\subsection{Data Collection and Analysis Procedures}
To answer our research questions, we prepared the NFRs from our study object and then performed a classification and analysis. The procedure was performed by the first two authors in a pair. Both have over three years of experience in requirements engineering research and model-based development research.

\subsubsection{Data Preparation}

The NFRs from our study objects differ in their level of abstraction, detail, and completeness. Therefore, we went through the set of NFRs and processed each of them in either one of the following ways:
\begin{itemize}
  \item {\bfseries Full interpretation:} We considered the NFR as it is.
  \item {\bfseries Sub interpretation:} We considered only a part of the NFR that we clearly identified as desired system property and disregarded the rest of the NFR (e.g., due to unnecessary\slash misleading information).
  \item {\bfseries Split requirement:} We split the NFR into a set of singular NFRs because the original NFR addressed more than one desired property of a system.
  \item {\bfseries Exclude from study:} We excluded the NFR if it was not in the scope of our study (e.g., process requirements), or if we were not able to understand the NFR due to missing or vague information.
\end{itemize}

In total, we excluded 56 requirements ($\approx16\%$) from
the study and considered 76 requirements ($\approx22\%$) only partially. We split 97
requirements ($\approx28\%$) into an overall of 337 requirements. Together with
the 117 requirements ($\approx34\%$) that we considered as they are, we ended
up with a set of 530 requirements that we used for our classification.

\subsubsection{Data Classification}
\begin{figure}
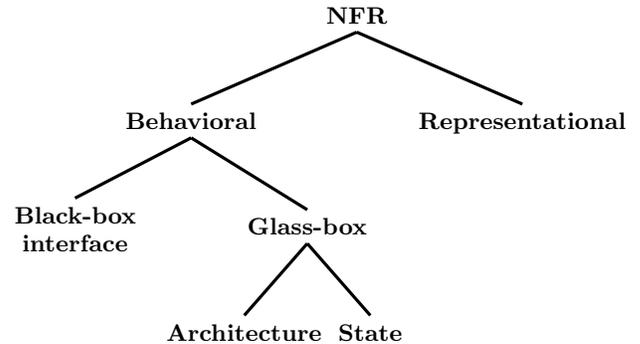


\begin{center}
\tikzset{edge from parent/.append style={very thick}}
\tikzset{level distance=40pt,sibling distance=0pt}
\Tree [.{\bfseries NFR}  [.{\bfseries Behavioral} [.{\bfseries \parbox{2cm}{\centering Black-box\\ interface}} ] [.{\bfseries Glass-box} {\bfseries Architecture} {\bfseries State} ] ] [.{\bfseries Representational} ] ]
\end{center}

  \caption{Classification of NFRs by means of the addressed \emph{system view}.  }
  \label{fig:decisiontree}
\end{figure}

We classified each of the 530 NFRs according to the following classification
schemes:
\begin{itemize}
\item{\bfseries NFR class:} We used the quality model for external and internal quality of the \ISO~\cite{iso9126} to assign a quality characteristic to each NFR (\emph{Func\-tionality--Suitability}, \emph{Reliability--Maturity}, \ldots ; see~\cite{iso9126} for details). In our study, the \ISO quality characteristics represent the NFR classes we consider.
\item{\bfseries System view:} We based our classification on Broy's structured views~\cite{Broy15NFR,Broy16Rethinking} to assign a {\itshape system view} to each NFR. As illustrated in Figure~\ref{fig:decisiontree}, structured views partition NFRs into \emph{representational} NFRs that refer to the way a system is syntactically or technically represented, described, structured, implemented, or executed (e.g., NFR of S4, Table~\ref{tbl:studyobjects}), and \emph{behavioral} NFRs that describe behavioral properties of a system. Behavioral NFRs are further partitioned into NFRs that describe \emph{black-box} behavior at the \emph{interface} of a system (e.g., NFR of S10, Table~\ref{tbl:studyobjects}) and NFRs that address a \emph{glass-box} view onto a system describing its \emph{architecture} (e.g., NFR of S8, Table~\ref{tbl:studyobjects}), or its \emph{state} behavior (e.g., NFR of S5, Table~\ref{tbl:studyobjects}).
\item{\bfseries Behavior theory:} Each \emph{behavioral} NFR uses a certain \emph{behavior theory} to express the desired properties of the system. We differentiate between the following classes of behavior theories for our classification:
\begin{itemize}
\item[\bfseries Syntactic] The NFR is expressed by a syntactic structure on which behavior can be described (e.g., NFR of S2, Table~\ref{tbl:studyobjects}). 
\item[\bfseries Logical] The NFR is expressed by a set of interaction patterns (e.g., NFR of S8, Table~\ref{tbl:studyobjects}).
\item[\bfseries Timed] The NFR is expressed by a set of interaction patterns with relation to time (e.g., NFR of S6, Table~\ref{tbl:studyobjects}).
\item[\bfseries Probabilistic] The NFR is expressed by probabilities for a set of interaction patterns (e.g., NFR of S1, Table~\ref{tbl:studyobjects}).
\item[\bfseries Timed and probabilistic] The NFR is expressed by probabilities for a set of interaction patterns with relation to time (e.g., NFR of S9, Table~\ref{tbl:studyobjects}).
\end{itemize}

\end{itemize}

To assess the feasibility and clarity of this classification scheme, we performed a pre-study on a subset of the NFRs (reported in our previously published material~\cite{Eckhardt15}). One result of this pre-study was a decision tree for the classification of NFRs. We created this tree to improve the  reproducibility of our classification (Figure~\ref{fig:decisiontree} shows a simplified version of the taxonomy on which the decision tree is based)\footnote{The decision tree can be found under:\\
\url{http://www4.in.tum.de/~eckharjo/DecisionTree.pdf}}. 

During the pre-study, we also recognized multiple occurrences of NFRs
following a common \emph{pattern}. For example, many specifications contained an
NFR following the pattern: {\itshape ``The system shall run\slash be installed on platform X''}. We identified a list of 13 of such patterns and assigned a common classification that we applied to all NFR instances following that pattern.\footnote{The complete list of patterns and the corresponding classification can be found under:\\
\url{http://www4.in.tum.de/~eckharjo/PatternList.pdf}}


\subsubsection{Data Analysis Procedures}
To answer RQ1, we analyzed the distribution of NFRs with respect to the \ISO quality characteristics. 
We provide two views onto this distribution. One detailed view that shows the distribution of NFRs with respect to all 27 quality characteristics contained in the standard and one coarse-grained view that shows the distribution of NFRs with respect to only 7 aggregated quality characteristics. In the aggregated quality characteristics, we subsumed low-level quality characteristics (such as {\itshape Functionality--Suitability} and {\itshape Functionality--Accuracy}) to their corresponding high-level quality class ({\itshape Functionality} in this case). We made one exception: we created for {\itshape Functionality--Security} an own class, as most other NFR classifications handle security separately. This results in the following aggregated list of quality characteristics: {\itshape Functionality}, {\itshape Usability}, {\itshape Reliability}, {\itshape Security}, {\itshape Efficiency}, {\itshape Maintainability}, and {\itshape Portability} (in the following we will refer to this list as \shortISO quality characteristics).

To answer RQ2, we analyzed the distribution of NFRs in the \shortISO quality characteristics with respect to the application domain of the corresponding system. 

To answer RQ3, we contrast the number of \emph{representational} NFRs with the number of \emph{behavioral} NFRs. To answer RQ4, we analyze the distribution of the behavioral NFRs with respect to {\itshape interface}, {\itshape architecture}, and {\itshape state behavior}. To answer RQ5, we analyze the distribution of the behavioral NFRs with respect to the \emph{behavior theory} used to express them. For each RQ, we present the results for the set of all requirements and structured according to the \shortISO quality characteristics.

\section{Study Results}
\label{results}
In the following, we report on the result for our research questions structured according to the research questions introduced in Section~\ref{studydesign}.

\subsection{Distribution of NFR Classes in Practice} 
\begin{table}
\scriptsize
\caption{Distribution of NFRs with respect to the \ISO quality characteristics}
\label{tbl:rq11}
\centering
\begin{tabularx}{\columnwidth}{lRR}\toprule
{\bfseries Quality characteristic}& {\bfseries count} & {\bfseries \%}\\ \midrule
Functionality - Suitability & 117 & 22.1\%\\              
Functionality - Security & 104  & 19.6\%\\
Reliability - Maturity & 40 & 7.5\%\\               
Usability - Operability & 40 & 7.5\%\\               
Efficiency - Time Behaviour & 37 & 7.0\%\\                     
Reliability - Reliability Compliance & 29  & 5.5\%\\                 
Efficiency - Resource Utilization & 21 & 4.0\%\\                 
Portability - Adaptability & 21 & 4.0\%\\
Portability - Installability & 18  & 3.4\%\\                
Maintainability - Changeability & 12  & 2.3\%\\
Reliability - Recoverability & 11 & 2.1\%\\
Functionality - Functionality Compliance & 10  & 1.9\%\\             
Usability - Learnability & 10 & 1.9\%\\
Functionality - Accuracy & 9  & 1.7\%\\
Usability - Usability Compliance & 9  & 1.7\%\\ 
Functionality -Interoperability & 8 & 1.5\%\\                     
Usability - Understandability & 8 & 1.5\%\\            
Maintainability - Analyzability & 7 & 1.3\%\\             
Reliability - Fault Tolerance  & 6  & 1.1\%\\                    
Maintainability - Stability & 4  & 0.8\%\\                  
Portability - Replaceability & 4  & 0.8\%\\
Portability - Co-Existence & 3 & 0.6\%\\               
Maintainability - Maintainability Compliance  & 1 & 0.2\%\\                
Usability - Attractiveness & 1 & 0.2\%\\        
\bottomrule           
\end{tabularx}
\end{table}

\begin{table}[h]
\centering
\scriptsize
\caption{Distribution of NFRs with respect to the \shortISO quality characteristics.}
\label{tbl:rq12}
\begin{tabularx}{\columnwidth}{lRR}\toprule
{\bfseries Quality characteristic}& {\bfseries count} & {\bfseries \%}\\ \midrule
Functionality  & 144 & 27.2\%\\     
Security & 104  &19.6\%\\      
Reliability & 86  &16.2\%\\        
Usability & 68  &12.8\%\\     
Efficiency & 58 &10.9\%\\ 
Portability & 46 &8.7\%\\
Maintainability & 24  &4.5\%\\     
\bottomrule
\end{tabularx}
\end{table}

\subsubsection*{RQ1: NFR Classes}

Table~\ref{tbl:rq11} shows the number (count) and percentage of NFRs (relative to the total number of NFRs) for each quality characteristic. Table~\ref{tbl:rq12} further shows the distribution with respect to the \shortISO quality characteristics. As shown in Table~\ref{tbl:rq11}, the two classes {\itshape Functionality--Suitability} and {\itshape Functionality--Security} stand out with in total 221 NFRs ($\approx$41.7\%). {\itshape Functionality--Suitability} is defined as {\itshape ``the capability of the software product to provide an appropriate set of functions for specified tasks and user objectives''}~\cite{iso9126}. This essentially corresponds to a classical understanding of a functional requirement. Furthermore, we classified up to 40 NFRs ($\approx$7.5\%) as {\itshape Reliability--Maturity}, {\itshape Usability--Operability}, or as {\itshape Efficiency--Time Behaviour}. 

In the aggregated results shown in Table~\ref{tbl:rq12}, one can see that the most common classification of NFRs is {\itshape Functionality} with around 27\%. 
Furthermore, around 20\% of all NFRs concern {\itshape Security}, 16\% concern {\itshape Reliability}, and 13\% concern {\itshape Usability}. {\itshape Efficiency} ($\approx$ 11\%), {\itshape Portability} ($\approx$ 9\%), and {\itshape Maintainability} ($\approx$ 5\%) occur only to a small extent in our data. 

\subsubsection*{RQ2: Relation to Application Domain}
\begin{figure}
\centering
  \includegraphics[width=\columnwidth]{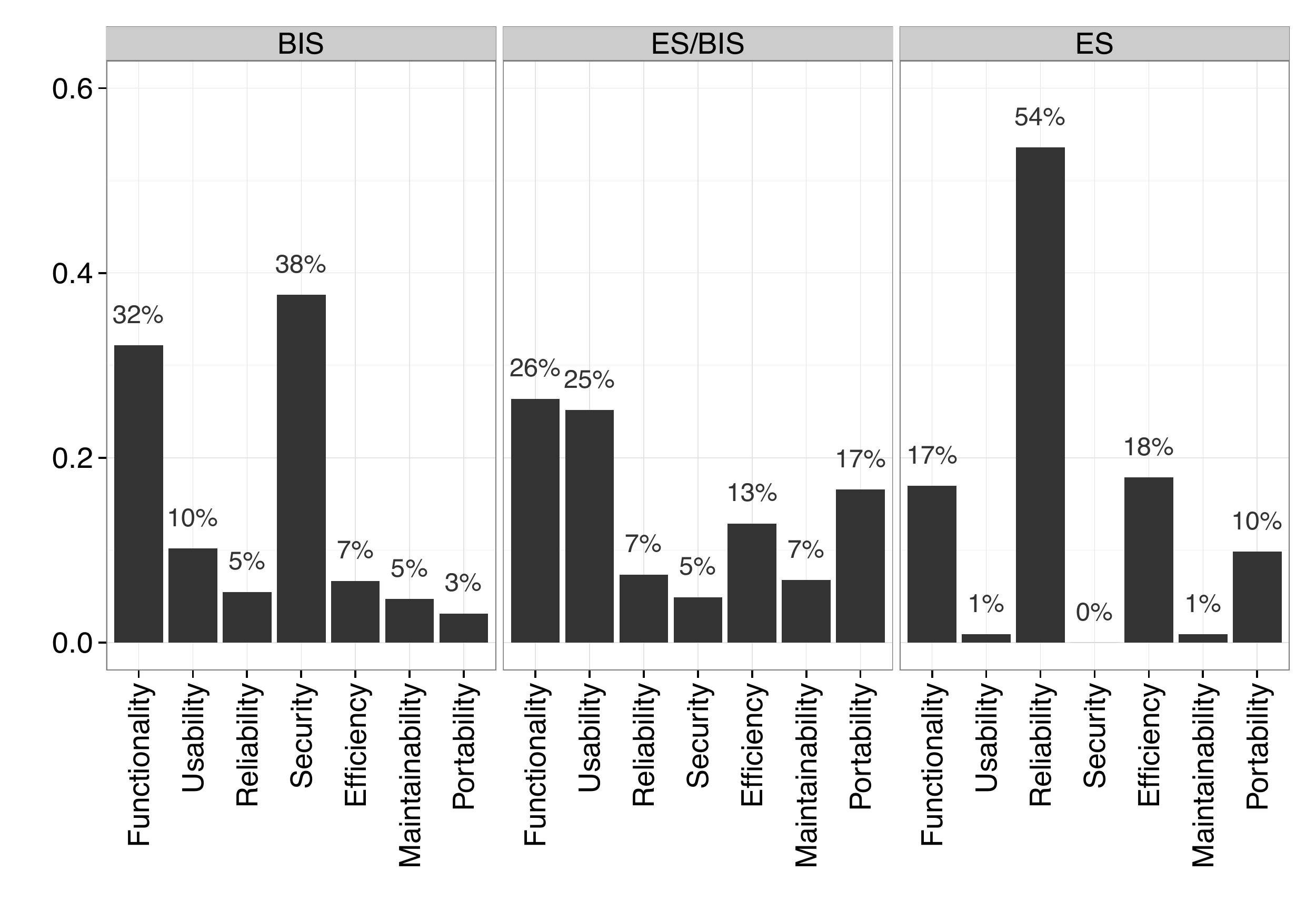}
  \caption{Relative distribution of NFRs over the \shortISO quality characteristics w.r.t. the application domain.}
  \label{fig:rq2}
\end{figure}

The results for RQ2 are given in Figure~\ref{fig:rq2} showing the distribution of NFR quality characteristics with respect to the application domain of the corresponding system ({\itshape Business Information System} (BIS), {\itshape Hybrid} (ES/BIS), or {\itshape Embedded System} (ES)). 

One can see a clear difference in the distribution of quality characteristics among the application domains. For example, for business information systems, we classified most NFRs as {\itshape Security} or {\itshape Functionality}, while for embedded systems, most NFRs are classified as {\itshape Reliability}. In hybrid systems, the distribution among the quality characteristics is more balanced compared with the other application domains. 

Although we expected to see different distributions of NFR classes between application domains, we were surprised by the extent of this difference. We see these results as a strong argument for domain-specific handling of NFRs. In Section~\ref{discussion}, we will discuss this in more detail.
\newpage

\subsection{Nature of NFR Classes} 

\subsubsection*{RQ3: Amount of NFRs Describing System Behavior}
\begin{figure}
\centering
\scriptsize
\begin{tabularx}{\columnwidth}{lRR}\toprule
{\bfseries Behavioral vs. Representational }& {\bfseries count} & {\bfseries \%}\\ \midrule
Behavioral  & 396 & 74.7\%\\     
\hspace{1em}-- Black-box  & 273 & 51.5\%\\     
\hspace{1em}-- Glass-box & 123  & 23.2\%\\      
Representational & 134  &25.3\%\\        
\bottomrule
\end{tabularx}

\vspace{1em}

 \includegraphics[width=\columnwidth]{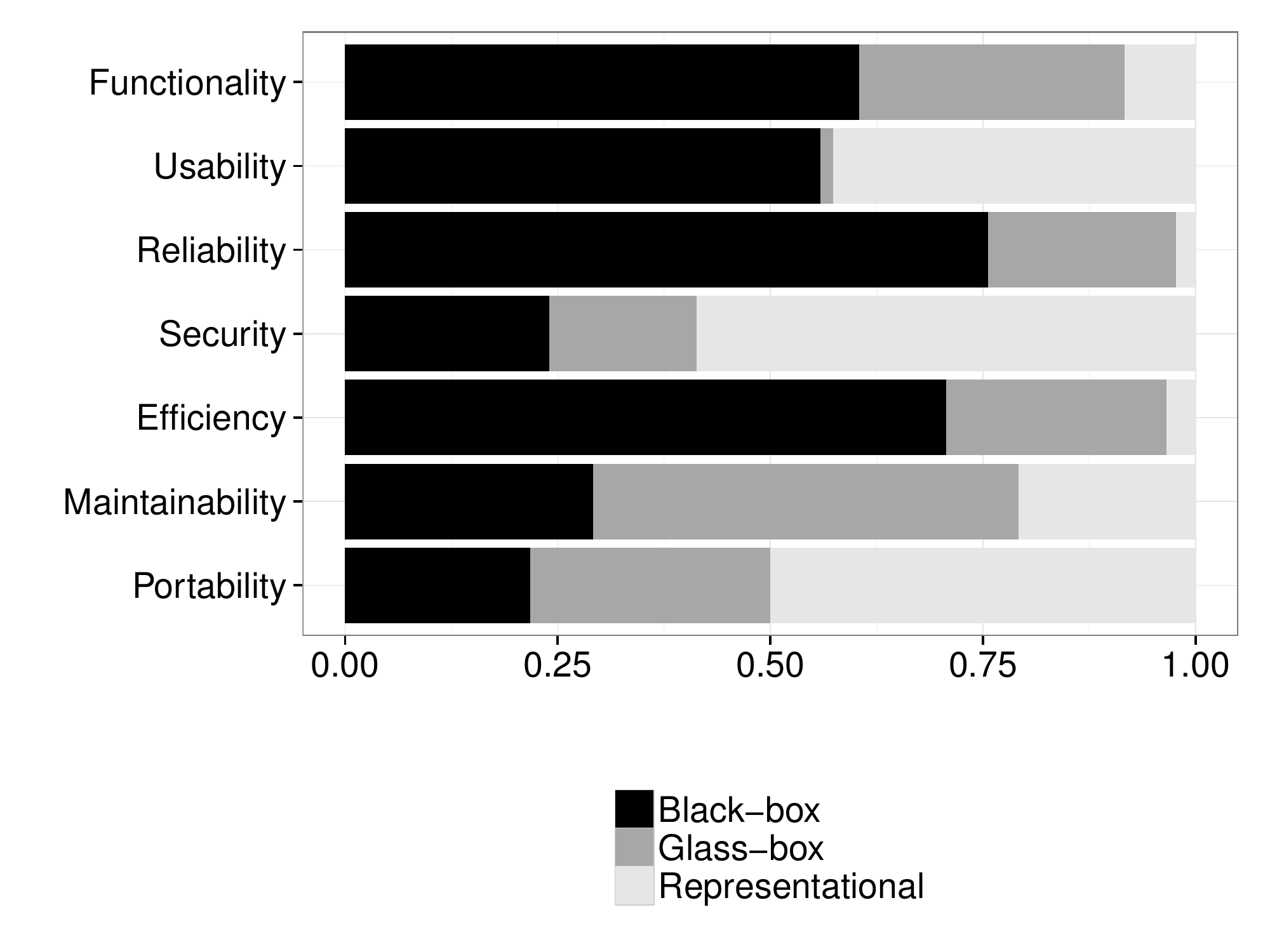}
 \caption{Distribution of \emph{behavioral} and \emph{representational} NFRs: \emph{black-box} (black), \emph{glass-box} (dark gray), and \emph{representational} (light gray).}%
 \label{fig:rq3}
\end{figure}

The results for RQ3 are shown in Figure~\ref{fig:rq3}. The table shows the distribution of \emph{behavioral} and \emph{representational} NFRs for all NFRs from our data set while the bar chart shows the distribution with respect to the \shortISO quality characteristics. More precisely, the bar chart shows the percentage of NFRs that we classified as {\itshape black-box} (black), {\itshape glass-box} (dark gray), or {\itshape representational} (light gray) within each \shortISO class.

\begin{center}
\noindent\fbox{\parbox{0.95\columnwidth}{
\begin{center}
\parbox{0.9\columnwidth}{
{\bfseries Quantitative results of RQ3:}\\
74.7\% of all NFRs describe behavior of the system (black-box or glass-box) while 25.3\% describe representational aspects.
}
\end{center}
}}
\end{center}
More than half of each of the NFRs in the {\itshape Functionality}, {\itshape Usability}, {\itshape Reliability}, and {\itshape Efficiency} classes describe {\itshape black-box} behavior defined over the interface of the system.  For example, most efficiency requirements describe desired or expected time intervals between events that are observable at the system interface. Reliability requirements often describe the observable reaction of the system at the interface if an error occurs within the system, such as {\itshape ``The [system] must have a mean time between failures greater than [x] h''}. 

The only class where the largest share of NFRs is classified as {\itshape glass-box} behavior is {\itshape Maintainability}. 
That means, maintainability requirements, if they consider system properties, often describe the desired internal structure or behavior within this structure (glass-box), as for example the requirement {\itshape ``The configuration [of the system] shall be independent from the system software and application software''}. However, a substantial amount of {\itshape glass-box} behavior can also be found in the {\itshape Functionality}, {\itshape Reliability}, {\itshape Security}, {\itshape Efficiency}, and {\itshape Portability} classes. Thus, NFRs within these classes also describe internal behavior, as for example the {\itshape Portability} requirement {\itshape ``The server software shall have the capability to run together with other applications on the same hardware whenever possible''}. 

Considering the amount of {\itshape representational} NFRs, one can see that the NFR classes {\itshape Usability}, {\itshape Portability}, and {\itshape Security} stand out. For usability and portability, this is as we expected. Usability requirements often describe representational aspects of the user interface with the goal to support the user in understanding and controlling a system, as for example the requirement {\itshape ``[The] GUI shall provide a common look and feel whenever possible''}. Portability requirements demand the system to be represented in a way that it fits a specified environment, as for example the requirement {\itshape ``The system shall run on platform X''}. However, for security, we did not expect such a high portion of \emph{representational} NFRs. Therefore, we analyzed these in detail and found that many representational NFRs in the {\itshape Security} class contain a reference to a standard. For example, we found a high number of NFRs stating, {\itshape ``The security class of the interface to system X with respect to data confidentiality is {\bfseries high}''}. Excluding those NFRs that reference standards from the results, around 54\% of security NFRs describe {\itshape black-box} behavior, 39\% describe {\itshape glass-box} behavior, and only 7\% describe {\itshape representational} aspects. This shows that some aspects of security are visible at the interface, as for example user authentication, and some aspects are internal to the system, as for example an encrypted communication within sub-systems.

Another point interesting to us was that none of the NFR classes is exclusively {\itshape black-box}, {\itshape glass-box}, or {\itshape representational}. For example, in the {\itshape Functionality} class, most NFRs describe {\itshape black-box} behavior. However, around 31\% of the NFRs describe glass-box behavior and 17\% describe representational aspects. This is because the {\itshape Functionality} class does not only include behavior over the interface, but also internal behavior like {\itshape``A system component shall save a user's edits whenever possible''}, and also representational aspects like {\itshape``The backup data must be stored according to [the company's] policies''}.

\subsubsection*{RQ4: Distribution of Behavioral NFRs w.r.t. System Views}
\begin{figure}
\centering
\scriptsize
\label{tbl:rq4}
\begin{tabularx}{\columnwidth}{lRR}\toprule
{\bfseries System view}& {\bfseries count} & {\bfseries \%}\\ \midrule
Interface  & 273 & 68.9\%\\     
Architecture & 85  & 21.5\%\\      
State & 38  &9.6\%\\    
\bottomrule 
\end{tabularx}

\vspace{1em}

  \includegraphics[width=\columnwidth]{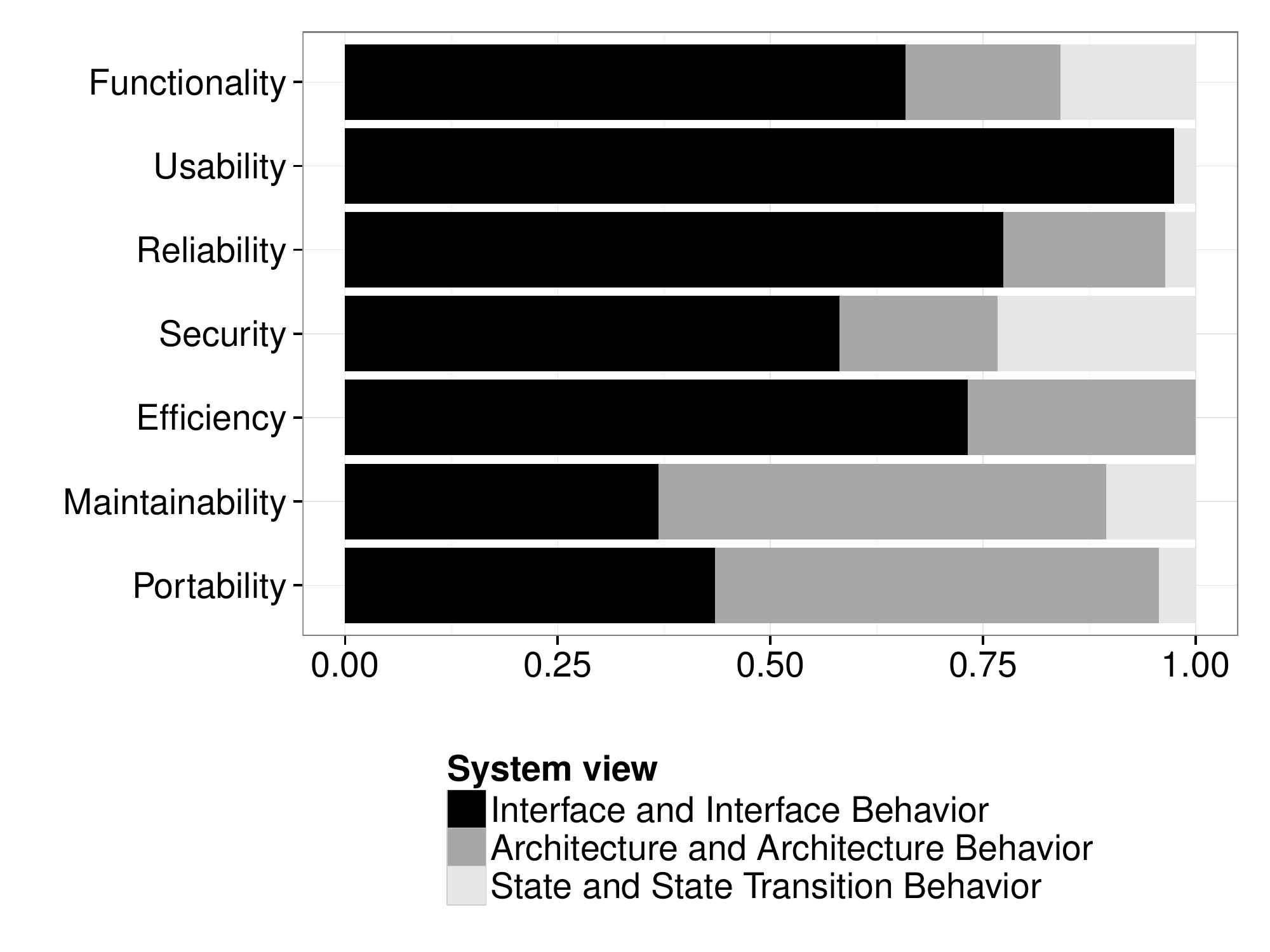}
 \caption{Distribution of \emph{behavioral} NFRs with respect to the \emph{system view} they address: \emph{interface} (black), \emph{architecture} (dark gray), and \emph{state} (light gray).}%
  \label{fig:rq4}
\end{figure}

The results for RQ4 are shown in Figure~\ref{fig:rq4}. The table shows the distribution of NFRs with respect to the {\itshape system view} they address while the bar chart shows this distribution with respect to the \shortISO quality characteristics. More precisely, the bar chart shows the percentage of NFRs that we classified as {\itshape interface} (black), {\itshape architecture} (dark gray), or {\itshape state} (light gray). For RQ4, we considered only \emph{behavioral} NFRs and neglected NFRs classified as \emph{representational}, as they do not describe behavior.

\begin{center}
\noindent\fbox{\parbox{0.95\columnwidth}{
\begin{center}
\parbox{0.9\columnwidth}{
{\bfseries Quantitative results of RQ4:}\\
68.9\% of all behavioral NFRs describe behavior over the interface of the system,  21.5\% describe architectural behavior, and 9.6\% describe behavior related to states of the system.
}
\end{center}
}}
\end{center}

We can see that in the {\itshape Functionality}, {\itshape Usability}, {\itshape Reliability}, {\itshape Security}, and {\itshape Efficiency} classes most behavioral NFRs are classified as {\itshape interface}. 
For the {\itshape Maintainability} and {\itshape Portability} classes, the most common classification is {\itshape architecture}. {\itshape Usability} is the only NFR class without any NFR classified as {\itshape architecture}.
We can further see that all NFR classes but {\itshape Efficiency} contain NFRs that describe state-related aspects, as for example the {\itshape Functionality} requirement {\itshape ``[The system] must ensure that submitted offers can neither be modified nor deleted''}. This shows that behavioral NFRs describe externally visible behavior but also behavior concerning the architecture (see structuring the functionality by functions~\cite{Broy10}) or state-related behavior (see operational states of a system~\cite{Vogelsang15}).
For example, in the {\itshape Security} class, there are NFRs that describe behavior over the interface like {\itshape ``There has to be an authentication mechanism''}, some NFRs describe architectural behavior like {\itshape ``[The system] must provide intrusion detection mechanisms''}, and some describe state-related aspects like {\itshape ``The password shall be valid for at most 30 days''}.

\subsubsection*{RQ5: Distribution of Behavioral NFRs w.r.t. Behavior Theories}
\begin{figure}
\centering
\scriptsize
\begin{tabularx}{\columnwidth}{lRR}\toprule
{\bfseries Behavior theory}& {\bfseries count} & {\bfseries \%}\\ \midrule
Syntactic & 47 & 11.9\%\\     
Logical & 277& 69.9\%\\     
Timed  & 54& 13.6\%\\     
Probabilistic & 7& 1.8\%\\     
Probabilistic \& Timed & 11& 2.8\%\\ 
\bottomrule    
\end{tabularx}

\vspace{1em}

  \includegraphics[width=\columnwidth]{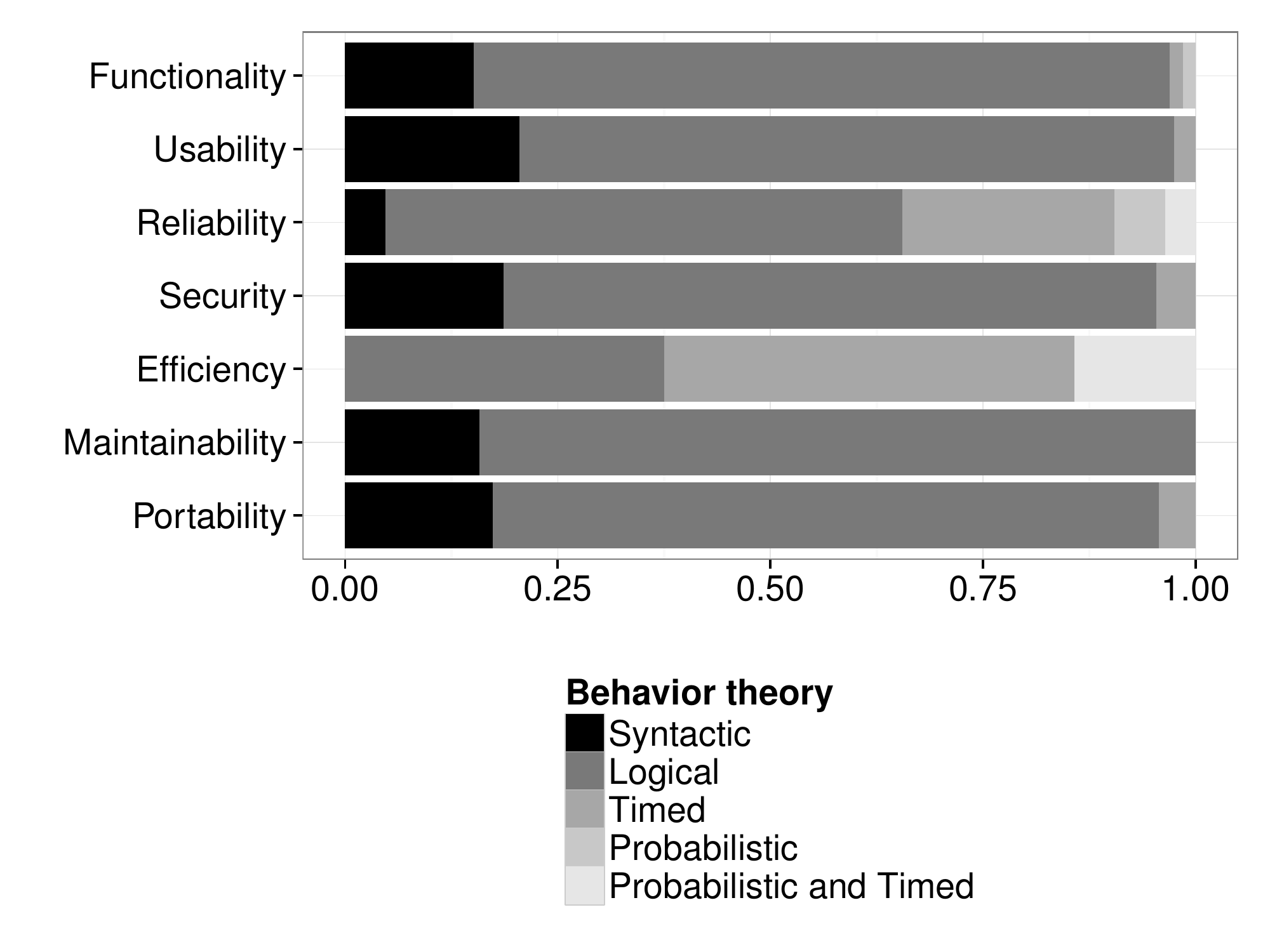}
  \caption{Relative distribution of behavioral NFRs with respect to their \emph{behavior theory}: {\itshape syntactic}, {\itshape logical}, {\itshape timed}, {\itshape probabilistic}, and {\itshape probabilistic and timed} (from black to white).}
  \label{fig:rq5}
\end{figure}

The results for RQ5 are shown in Figure~\ref{fig:rq5}. The table shows the distribution of NFRs with respect to the {\itshape behavior theory} they use and the bar chart shows this distribution with respect to the \shortISO quality characteristics. More precisely, the figure shows the percentage of NFRs that we classified as {\itshape syntactic}, {\itshape logical}, {\itshape timed}, {\itshape probabilistic}, or {\itshape probabilistic and timed} (from black to white). For RQ5, we considered only \emph{behavioral} NFRs and neglected NFRs classified as \emph{representational} as they do not describe behavior.

\begin{center}
\noindent\fbox{\parbox{0.95\columnwidth}{
\begin{center}
\parbox{0.9\columnwidth}{
{\bfseries Quantitative results of RQ5:}\\
Most behavioral NFRs are logical (69.9\%), 18.2\% are timed and\slash or probabilistic, and only 11.9\% are syntactic.
}
\end{center}
}}
\end{center}

Over all NFR classes, most NFRs are {\itshape logical} (around 69.9\%), while 13.6\% are {\itshape timed}, 11.9\% are {\itshape syntactic}, 2.8\% are {\itshape probabilistic and timed}, and 1.8\% are {\itshape probabilistic}. Most {\itshape timed} and also {\itshape probabilistic and timed} NFRs belong to the {\itshape Efficiency} class. Moreover, the {\itshape Reliability} class stands out, as it also contains many {\itshape timed}, {\itshape probabilistic}, and {\itshape timed and probabilistic} NFRs.

\subsection{Summary of Results} 

\begin{figure}[h!]
\centering
  \includegraphics[width=\columnwidth]{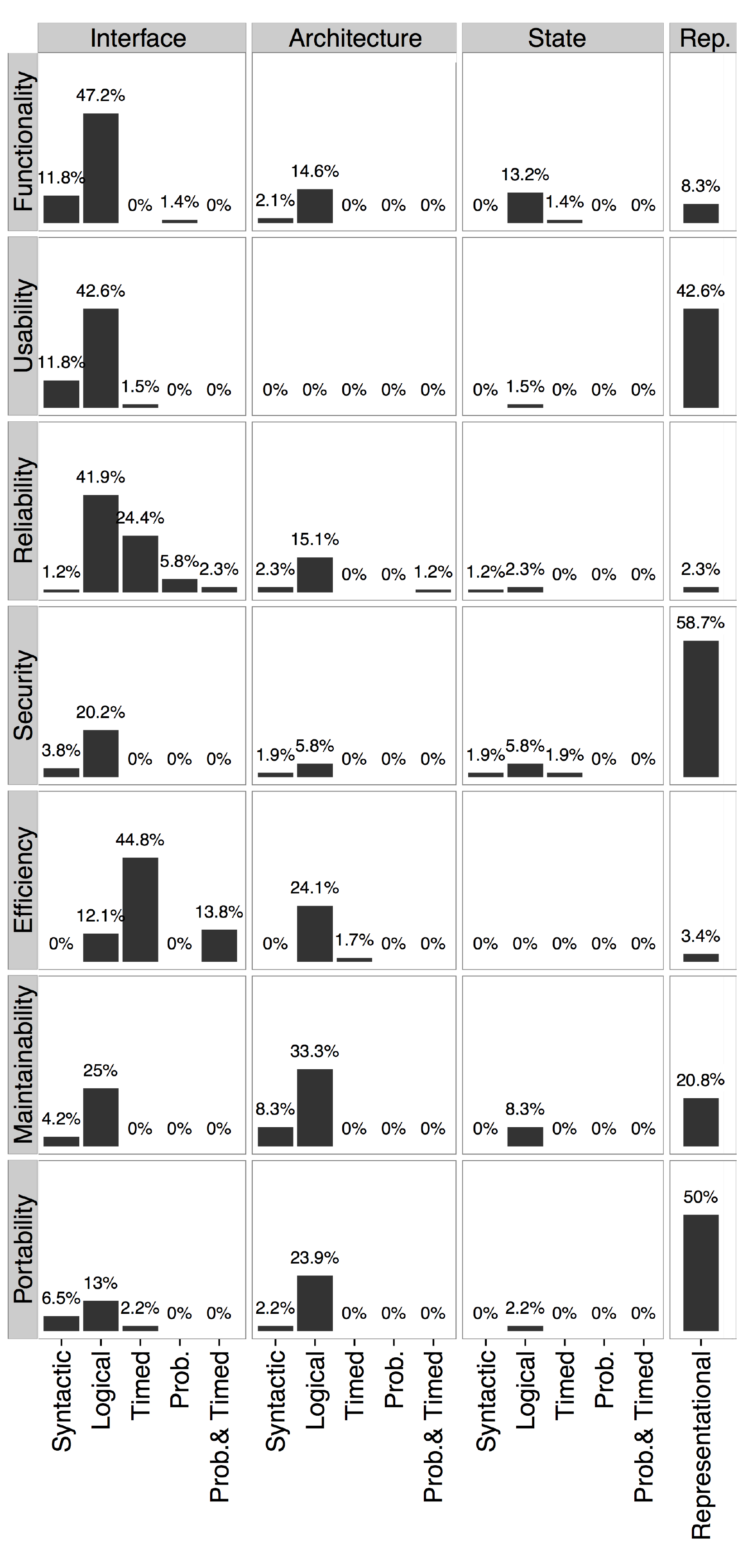}
  \caption{Relative distribution of NFRs within the NFR classes w.r.t.\ the addressed \emph{system view} and the used \emph{behavior theory}.
  }
  \label{fig:discussion}
\end{figure}

Figure~\ref{fig:discussion} provides a consolidated quantified view on our overall results. 

It shows the distribution of NFRs among the NFR classes with respect to the addressed \emph{system view} and the used {\itshape behavior theory}. The figure shows a table with one diagram per cell; the rows display the NFR classes and the columns display the addressed {\itshape system view} and an additional column for the \emph{representational} NFRs. Within each cell, the relative distribution per {\itshape behavior theory} is shown (relative per NFR class, i.e., the values in all cells of each row sum up to 100\%).

In conclusion, most NFRs address interface behavior, mostly expressed by logical or timed assertions. The NFR classes \emph{Usability}, \emph{Security}, and \emph{Portability} include, in contrast to the other classes, a high portion of \emph{representational} NFRs. Furthermore, all classes but \emph{Usability} contain architectural aspects (see column {\itshape Architecture}), while the highest percentage of those NFRs are in the {\itshape Maintainability} class.

\section{Threats to Validity}
\label{threats}

In the following, we discuss the threats to validity and mitigation measures we took. We discuss them along the different problems as they arose during our work.

\subsection{Data Representativeness Problem} Inherent to the nature of our study is the data representativeness on which we built our analysis. The concerns range from the representativeness of the way the NFRs are
specified to the completeness of the data as it currently covers only the particularities of selected industrial contexts. We cannot mitigate this threat but consider our data set large enough to allow us to draw first conclusions on the state of the practice. The relation to existing evidence (see Section~\ref{sect:reltoevidence}) additionally strengthens our conclusions.

\subsection{NFR Selection Problem} We collected only those requirements explicitly labeled as non-functional or quality. With this selection procedure, some relevant NFRs might have been missed or irrelevant ones might have been included. To address this problem, we plan to perform the classification on the whole data set as future work, including functional and non-functional requirements. 

\subsection{Preparation Problem} In our data preparation phase, we excluded NFRs from the study if they were not in scope of our study (e.g., process requirements) or if we were not able to understand them (due to missing or vague information). This exclusion process could threaten the overall conclusion validity, but as we excluded only about 16\%, we do not consider this as a major threat.

\subsection{Classification Problem} Prior to our study, we performed a pre-study with several independent classification rounds~\cite{Eckhardt15}. The inter-rater agreement between the independent raters was, however, so low that we had to conclude that the classification dimensions are not clear enough. To resolve this issue, we performed several refinements of the classification and created a decision tree and a pattern catalogue that supports the classification process~\cite{Eckhardt15}. In the end, we did the classification in a pair of researchers and individually discussed each NFR.

\subsection{Representation Problem} Although classifying in a pair of researchers, we still faced the \emph{representation problem} discussed by Glinz~\cite{glinz2007non}, which threatens the internal validity. If an NFR stated {\itshape ``The system shall authenticate the user''}, we classified it as {\itshape black-box interface}, and {\itshape logical} as it describes a black-box behavior over the interface. However, if an NFR stated {\itshape ``The system shall contain an authentication component''}, we classified it as {\itshape glass-box architecture} and {\itshape logical} as it requires an {\itshape internal} component for authentication. 

\subsection{Contextualization Problem} We consider the reliability of our conclusions to be very much
dependent on the possibility to reproduce the results, which in turn is dependent on the clearness of the context information. The latter, however, is strongly limited by NDAs that too often prevent providing full disclosure of the contexts and even the project characteristics. To mitigate this threat, we anonymized the data as much as possible and disclosed all information possible within the limits of our non-disclosure agreements.

\section{Discussion}
\label{discussion}
Based on the results, we identified a set of insights which we discuss in the following paragraphs. 

\paragraph*{NFRs are not non-functional}
It is commonly acknowledged that functional requirements describe logical behavior over the interface of the system. From a broader view, one could even say that functional requirements describe any kind of behavior over the interface of the system, including timing and\slash or probabilistic behavior. From this perspective, we conclude that many of those NFRs that address system properties describe the same type of behavior as functional requirements do (see column {\itshape Interface} in Figure~\ref{fig:discussion}). This is true for almost all NFR classes we analyzed; even for NFR classes which are sometimes called \emph{internal} quality attributes (e.g., portability or maintainability)~\cite{mcconnell2004code}. Hence, we argue that---at least based on our data---most ``non-functional'' requirements describe functional aspects of a system and are, thus, basically \emph{not} non-functional. From a practical point of view, this means that most NFRs can be elicited, specified, and analyzed like functional requirements. For example, NFRs classified as black-box interface requirements, are candidates for system tests. In our data set, system test cases could have been specified for almost 51.5\% of the NFRs.

\paragraph*{Functional requirements are often labeled as NFRs} 
Moreover, functional requirements in the classical understanding were often labeled as NFRs in our examined specifications. We classified 22.1\% of our overall NFR population as {\itshape Functionality--Suitability}, which is a quality characteristic that addresses the functionality of a system (\emph{``The capability of the software product to provide an appropriate set of functions for specified tasks and user objectives''}~\cite{iso9126}). Given that NFRs are usually not tested and analyzed as thoroughly as functional requirements~\cite{Ameller12,borg2003bad,svensson2009quality}, this means that, one out of five NFRs in our data set elude a thorough analysis process just because they are labeled as NFRs. 

\paragraph*{NFRs are often specified by reference to standards}
As already indicated within our results for RQ3, we realized that several examined NFRs describe requirements by pointing to a standard (e.g., company style guides or safety standards). More specifically, 68 of 530 NFRs ($\approx13\%$) contained references to standards. We classified these as \emph{representational} since we were not able to access these standards due to availability and time constraints. However, these standard-referencing NFRs might be interesting to explore in future investigations. On the one hand, they allow a concise specification; on the other hand, they introduce much implicitly necessary knowledge and assume that the reader of the specification has knowledge about and access to those standards. 

\paragraph*{Only few NFRs deal with architectural aspects}
While in literature the relation of NFRs to architecture and architectural constraints of a system is often emphasized~\cite{chung2012non,Pohl2010RE,zhu2007uml}, the NFRs of our sample dealt with architecture only to a small degree (see column {\itshape Architecture} in Figure~\ref{fig:discussion}). Only for {\itshape Efficiency}, {\itshape Maintainability}, and {\itshape Portability}, roughly one quarter of the NFRs considered architectural aspects of a system. Following this, we argue that---at least based on our data---only few NFRs actually describe architectural aspects of a system. It is an interesting point for future research to mirror our findings with the notion of \emph{architecturally significant requirements (ASRs)}~\cite{Chen13}. ASRs are those requirements which have a measurable impact on a software system's architecture. They are often difficult to define and articulate, tend to be expressed vaguely, are often initially neglected, tend to be hidden within other requirements, and are subjective, variable, and situational~\cite{Chen13}. Certainly, all NFRs that we classified as addressing the system view \emph{architecture} can be considered as ASRs. However, also NFRs that we classified as addressing the system view \emph{interface} or \emph{state} may have an impact on the architecture, as for example the requirement {\itshape ``the system should provide five nines (99.999 percent) availability''}. The difference is that NFRs addressing the system view \emph{architecture} make the impact on the architecture explicit. For other NFRs, an architect needs to decide whether they are ASRs or not. 

\paragraph*{No NFR class is uniquely affiliated with only one behavior characteristic}
Our analysis shows that none of the considered NFR classes is characterized by only one specific {\itshape system view} or {\itshape behavior theory}. Accordingly, most NFR classes contain representational and behavioral NFRs which address all {\itshape system views} and using all {\itshape behavior theories}. While a classification of NFRs according to quality characteristics may be helpful to express the intent of an NFR, the quality characteristic should not determinet how an NFR is specified, implemented, or tested. This decision should rather be made based on the addressed {\itshape system view} and the {\itshape behavior theory} used to express the NFR.

\paragraph*{The application domain influences NFRs}
As our results indicate, the application domain of the corresponding system influences the relevancy of NFR classes. We therefore conclude that specification and analysis procedures should be customized for different application domains. For example, in the embedded systems domain, the need for probabilistic analysis techniques is stonger compared with the business information systems domain due to the larger amount of reliability NFRs that are often described in a probabilistic manner. On the other hand, for business information systems, we should support specification techniques that integrate functional requirements and behavioral NFRs, since around 70\% of the NFRs from BISs were classified as functionality or security (excluding the NFRs referencing standards from the security class, most NFRs in the security class concern the interface), which describe logical interface behavior to a large extent.

\paragraph*{NFRs are specified at different levels of abstraction}
In our data set, we found NFRs at different levels of abstraction varying in their degree of detail and completeness. NFRs ranged from high-level goal descriptions like {\itshape ``The availability shall not be less than [x]\%''} 
to very concrete and detailed descriptions of behavior like {\itshape ``The delay between passing a [message] and decoding the first loop message shall be $\leq$ [x] seconds''}. This is in tune with the view of Pohl~\cite{Pohl2010RE}; He states that non-functional requirements are underspecified functional requirements. In a development process, high-level NFRs need to be refined to detailed functional requirements. To make this refinement explicit, we need an approach for relating high-level NFRs (or quality goals) to low-level functional requirements. A first approach in this direction is proposed by Broy in his recent work~\cite{Broy16Rethinking}.

\section{Conclusions}
\label{conclusion}

In this paper, we reported on a study where we analyzed and classified 530 NFRs extracted from 11 industrial requirements specifications with respect to their kind and their relation to {\itshape system views} and {\itshape behavior theories}. Our goal was to gain a better understanding on the nature of system-specific NFRs, i.e., those NFRs that address system properties. We were able to show, for example, that most of the NFRs in our sample actually describe black-box interface behavior. Our overall conclusion is that NFRs are in their nature \emph{not} non-functional. Therefore, we argue that many so-called NFRs can be handled similarly to functional requirements and, thus, can be integrated into a seamless software engineering development process. 

\subsection{Relation to Existing Evidence}
\label{sect:reltoevidence}

Our results show various relations to existing evidence. For instance, during our classification, we faced all three problems described by Glinz~\cite{glinz2007non}. We also experienced same or similar terminological confusions on NFRs as reported by Ameller~et~al.~\cite{Ameller12}. In particular, we found that categories such as {\itshape availability} were often misinterpreted in the documents and used in different ways, e.g., as {\itshape performance}. Furthermore, they found that the four NFR classes most important to software architects were {\itshape performance}, {\itshape usability}, {\itshape security}, and {\itshape availability} (in that order). We could support their results via quantitative results: Their four NFR classes are in our list of the top four NFR classes (in a different order). Finally, our results also resemble the results of Mairiza~et~al.~\cite{mairiza2010investigation} with respect to the five most frequently mentioned NFR classes in literature.

Apart from supporting existing evidence, we provide first empirical evidence on what non-functional (system) requirements  are in their nature by analyzing and classifying them with respect to various facets. Summarizing our findings, we conclude that most so-called ``non-functional'' requirements in our sample describe functional aspects of a system and are, thus, essentially \emph{not} non-functional.

\subsection{Impact\slash Implications}
\label{sect:impact}
Our results strengthen our confidence that many requirements that are currently classified as NFRs in practice can be handled equally to functional requirements, which has both a strong theoretical and practical impact. 
Existing literature (e.g.,~\cite{Ameller12,borg2003bad,chung1995dealing,svensson2009quality}) indicates that the development process for a requirement differs depending on whether it is classified as ``non-functional'' or ``functional''. In contrast to functional requirements, requirements classified as NFR are often neglected and properties like testability are not supported. In industrial collaborations, we have also seen that NFRs and functional requirements were documented in separate documents, which has led to failing acceptance tests performed by an external company. Our results suggest that this separation is artificial to a large extent. We argue that treating NFRs the same as functional requirements would have major consequences for the software engineering process. However, there are currently no empirical studies that investigate this argument in detail.

A long-term vision that emerges from our results is that we might be able to better integrate NFRs into a holistic software development process in the future. Such an integration would yield, for instance, seamless modeling of all properties associated with a system---no matter if they are functional or non-functional in a classical understanding. The benefits of such an integration include that NFRs would not be neglected during development activities, as it is too often current state of practice; from an improvement in the traceability of requirements over an improvement of possibilities for progress control to an improvement of validation and verification.

\subsection{Future Work}
Our analysis is based on an inherently incomplete set of requirements specifications gathered from practical environments. Hence, our study can be considered as a first attempt to improve the understanding on the nature of NFRs from a practical perspective. This has certain implications on the validity of our results (see Section~\ref{threats}). However, they still provide a suitable basis to draw first conclusions, which need to be strengthened via additional studies; for instance, by increasing the sample size, by taking into account further application domains, but also by including functional requirements into the analysis. So far, our results are suitable to trigger critical, yet important discussions within the community.

We are planning three concrete next steps based on our data set: First, we will include the remaining 1495 functional requirements (the ones not labeled as ``non-functional'' or quality attribute) in our study. 
Second, we are planning to advance the integration of NFRs into software development by providing specification blueprints (based on an integrated model) for practitioners. Third, as discussed in Section~\ref{sect:impact}, we will investigate the consequences of labeling a requirement as ``NFR'' for the development process. We expect to find consequences for how requirements labeled as NFRs are tested 
or when they are considered in the development process.

\section*{Acknowledgements}
We would like to thank M. Broy, E. J\"urgens, M. Junker, B. Penzenstadler, and S. Smith-Eckhardt for their helpful comments on earlier versions of this work. Furthermore, we thank the anonymous reviewers for their detailed feedback on our work.

\bibliographystyle{abbrv}
\bibliography{nfrstudy}

\begin{thebibliography}{10}

\bibitem{Ameller12}
D.~Ameller, C.~Ayala, J.~Cabot, and X.~Franch.
\newblock How do software architects consider non-functional requirements: An
  exploratory study.
\newblock In {\em Proc. of the 20th IEEE International Requirements Engineering
  Conference (RE)}, 2012.

\bibitem{ameller2010dealing}
D.~Ameller, X.~Franch, and J.~Cabot.
\newblock Dealing with non-functional requirements in model-driven development.
\newblock In {\em Proc. of the 18th IEEE International Requirements Engineering
  Conference (RE)}, 2010.

\bibitem{borg2003bad}
A.~Borg, A.~Yong, P.~Carlshamre, and K.~Sandahl.
\newblock The bad conscience of requirements engineering: an investigation in
  real-world treatment of non-functional requirements.
\newblock In {\em Proc. of the 3rd Conference on Software Engineering Research
  and Practice in Sweden (SERPS)}, 2003.

\bibitem{Broy10}
M.~Broy.
\newblock Multifunctional software systems: {S}tructured modeling and
  specification of functional requirements.
\newblock {\em Science of Computer Programming}, 75(12), 2010.

\bibitem{Broy15NFR}
M.~Broy.
\newblock Rethinking nonfunctional software requirements.
\newblock {\em {IEEE} Computer}, 48(5), 2015.

\bibitem{Broy16Rethinking}
M.~Broy.
\newblock Rethinking nonfunctional software requirements: {A} novel approach
  categorizing system and software requirements.
\newblock In M.~Hinchey, editor, {\em Software Technology: 10 Years of
  Innovation in {IEEE} Computer}. John Wiley \& Sons\slash IEEE Press, 2016.

\bibitem{Chen13}
L.~Chen, M.~Ali~Babar, and B.~Nuseibeh.
\newblock Characterizing architecturally significant requirements.
\newblock {\em IEEE Software}, 30(2), 2013.

\bibitem{chung2009non}
L.~Chung and J.~C.~S. do~Prado~Leite.
\newblock On non-functional requirements in software engineering.
\newblock In {\em Conceptual modeling: {F}oundations and applications}, volume
  5600 of {\em Lecture Notes in Computer Science}. Springer, 2009.

\bibitem{chung1995dealing}
L.~Chung and B.~A. Nixon.
\newblock Dealing with non-functional requirements: three experimental studies
  of a process-oriented approach.
\newblock In {\em Proc. of the 17th International Conference on Software
  Engineering (ICSE)}, 1995.

\bibitem{chung2012non}
L.~Chung, B.~A. Nixon, E.~Yu, and J.~Mylopoulos.
\newblock {\em Non-functional requirements in software engineering}, volume~5.
\newblock Springer Science \& Business Media, 2012.

\bibitem{damm2005boosting}
W.~Damm, A.~Votintseva, A.~Metzner, B.~Josko, T.~Peikenkamp, and E.~B{\"o}de.
\newblock Boosting re-use of embedded automotive applications through rich
  components.
\newblock In {\em Proc. of the Workshop on Foundations of Interface
  Technologies (FIT)}, 2005.

\bibitem{Eckhardt15}
J.~Eckhardt, D.~M{\'e}ndez~Fern{\'a}ndez, and A.~Vogelsang.
\newblock How to specify non-functional requirements to support seamless
  modeling? {A} study design and preliminary results.
\newblock In {\em Proc. of the 9th International Symposium on Empirical
  Software Engineering and Measurement (ESEM)}, 2015.

\bibitem{fatwanto2008analysis}
A.~Fatwanto and C.~Boughton.
\newblock Analysis, specification and modeling of non-functional requirements
  for translative model-driven development.
\newblock In {\em Proc. of the 2008 International Conference on Computational
  Intelligence and Security (CIS)}, 2008.

\bibitem{glinz2007non}
M.~Glinz.
\newblock On non-functional requirements.
\newblock In {\em Proc. of the 15th IEEE International Requirements Engineering
  Conference (RE)}, 2007.

\bibitem{gonczy2009model}
L.~G{\"o}nczy, Z.~D{\'e}ri, and D.~Varr{\'o}.
\newblock Model transformations for performability analysis of service
  configurations.
\newblock In {\em Models in Software Engineering}, volume 5421 of {\em Lecture
  Notes in Computer Science}. Springer, 2009.

\bibitem{iso9126}
{ISO/IEC}.
\newblock Software engineering -- {P}roduct quality.
\newblock ISO/IEC 9126, International Organization for Standardization, Geneva,
  Switzerland, 2001.

\bibitem{Junker12}
M.~Junker and P.~Neubeck.
\newblock A rigorous approach to availability modeling.
\newblock In {\em Proc. of the 4th International Workshop on Modeling in
  Software Engineering (MiSE)}, 2012.

\bibitem{kugele2008optimizing}
S.~Kugele, W.~Haberl, M.~Tautschnig, and M.~Wechs.
\newblock Optimizing automatic deployment using non-functional requirement
  annotations.
\newblock In {\em Leveraging Applications of Formal Methods, Verification and
  Validation}, volume~17 of {\em Communications in Computer and Information
  Science}. Springer, 2008.

\bibitem{mairiza2010investigation}
D.~Mairiza, D.~Zowghi, and N.~Nurmuliani.
\newblock An investigation into the notion of non-functional requirements.
\newblock In {\em Proc. of the 25th ACM Symposium on Applied Computing (SAC)},
  2010.

\bibitem{mcconnell2004code}
S.~McConnell.
\newblock {\em Code complete}.
\newblock Pearson Education, 2004.

\bibitem{molina2009integrating}
F.~Molina and A.~Toval.
\newblock Integrating usability requirements that can be evaluated in design
  time into model driven engineering of web information systems.
\newblock {\em Advances in Engineering Software}, 40(12), 2009.

\bibitem{Mylopoulos92}
J.~Mylopoulos, L.~Chung, and B.~Nixon.
\newblock Representing and using nonfunctional requirements: a process-oriented
  approach.
\newblock {\em IEEE Transactions on Software Engineering}, 18(6), 1992.

\bibitem{Pohl2010RE}
K.~Pohl.
\newblock {\em Requirements Engineering: Fundamentals, Principles, and
  Techniques}.
\newblock Springer, 1st edition, 2010.

\bibitem{robertson2012mastering}
S.~Robertson and J.~Robertson.
\newblock {\em Mastering the requirements process: {G}etting requirements
  right}.
\newblock Addison-wesley, 2012.

\bibitem{Sommerville97}
I.~Sommerville and P.~Sawyer.
\newblock {\em Requirements engineering: a good practice guide}.
\newblock John Wiley \& Sons, Inc., 1997.

\bibitem{svensson2009quality}
R.~B. Svensson, T.~Gorschek, and B.~Regnell.
\newblock Quality requirements in practice: An interview study in requirements
  engineering for embedded systems.
\newblock In {\em Requirements Engineering: Foundation for Software Quality},
  volume 5512 of {\em Lecture Notes in Computer Science}. Springer, 2009.

\bibitem{Vogelsang15}
A.~Vogelsang, H.~Femmer, and C.~Winkler.
\newblock Systematic elicitation of mode models for multifunctional systems.
\newblock In {\em Proc. of the 23rd IEEE International Requirements Engineering
  Conference (RE)}, 2015.

\bibitem{wada2010model}
H.~Wada, J.~Suzuki, and K.~Oba.
\newblock A model-driven development framework for non-functional aspects in
  service oriented architecture.
\newblock {\em Web Services Research for Emerging Applications: Discoveries and
  Trends}, 2010.

\bibitem{Wagner12}
S.~Wagner, K.~Lochmann, L.~Heinemann, M.~Kl\"{a}s, A.~Trendowicz,
  R.~Pl\"{o}sch, A.~Seidl, A.~Goeb, and J.~Streit.
\newblock The quamoco product quality modelling and assessment approach.
\newblock In {\em Proc. of the 34th International Conference on Software
  Engineering (ICSE)}, 2012.

\bibitem{zhu2007uml}
L.~Zhu and I.~Gorton.
\newblock {UML} profiles for design decisions and non-functional requirements.
\newblock In {\em Proc. of the 2nd Workshop on Sharing and Reusing
  Architectural Knowledge Architecture, Rationale, and Design intent
  (SHARK-ADI)}, 2007.

\bibitem{zhu2009model}
L.~Zhu and Y.~Liu.
\newblock Model driven development with non-functional aspects.
\newblock In {\em ICSE Workshop on Aspect-Oriented Requirements Engineering and
  Architecture Design (EA)}, 2009.

\end{thebibliography}
\end{document}